\documentclass[journal]{IEEEtran}
\ifCLASSINFOpdf
\else
\fi
\usepackage{url}
\usepackage{cite}
\usepackage{amsmath,amssymb,amsfonts}
\usepackage{graphicx}
\usepackage{textcomp}
\usepackage{multicol}
\usepackage{multirow}
\usepackage{algorithm}
\usepackage[noend]{algpseudocode}
\usepackage{tabularx}
\usepackage{adjustbox}  
\usepackage{dblfloatfix}
\usepackage[para]{threeparttable}
\usepackage{subfigure}
\usepackage{caption}
\usepackage{comment}
\usepackage{fancyhdr}
\usepackage{mathtools}

\def\sgn{{\rm sgn}}

\fancypagestyle{github}{\fancyhf{}\fancyfoot[L]{$^*$https://github.com/aimlab-wustl/GiniSVMMicro}}

\begin{document}
%
\title{In-filter Computing For Designing Ultra-light Acoustic Pattern Recognizers}
%
%
%

\author{Abhishek~Ramdas~Nair,
        Shantanu~Chakrabartty,
        and~Chetan~Singh~Thakur
\thanks{Abhishek Ramdas Nair and Chetan Singh Thakur are with the Department
of Electronic Systems Engineering, Indian Institute of Science, Bangalore,
KA, 560012 INDIA e-mail: (abhisheknair@iisc.ac.in, csthakur@iisc.ac.in).}
\thanks{Shantanu Chakrabartty is with Department of Electrical and Systems Engineering, Washington University in St. Louis,USA, 63130 e-mail: (shantanu@wustl.edu).}
\thanks}

%
%

\markboth{}%
{A.R.Nair \MakeLowercase{\textit{et al.}}: In-filter Computing For Designing Ultra-light Acoustic Pattern Recognizers}
%



\maketitle

\begin{abstract}
We present a novel in-filter computing framework that can be used for designing ultra-light acoustic classifiers for use in smart internet-of-things (IoTs). Unlike a conventional acoustic pattern recognizer, where the feature extraction and classification are designed independently, the proposed architecture integrates the convolution and nonlinear filtering operations directly into the kernels of a Support Vector Machine (SVM). The result of this integration is a template-based SVM whose memory and computational footprint (training and inference) is light enough to be implemented on an FPGA-based IoT platform. While the proposed in-filter computing framework is general enough, in this paper, we demonstrate this concept using a Cascade of Asymmetric Resonator with Inner Hair Cells (CAR-IHC) based acoustic feature extraction algorithm. The complete system has been optimized using time-multiplexing and parallel-pipeline techniques for a Xilinx Spartan 7 series Field Programmable Gate Array (FPGA). We show that the system can achieve robust classification performance on benchmark sound recognition tasks using only \texttildelow 1.5k Look-Up Tables (LUTs) and \texttildelow 2.8k Flip-Flops (FFs), a significant improvement over other approaches.
\end{abstract}


%
\IEEEpeerreviewmaketitle

\section{Introduction}
%
%
%
%

\begin{figure}[htbp]
\centerline{\includegraphics[page=1,scale=0.32,trim=3 3 3 3,clip]{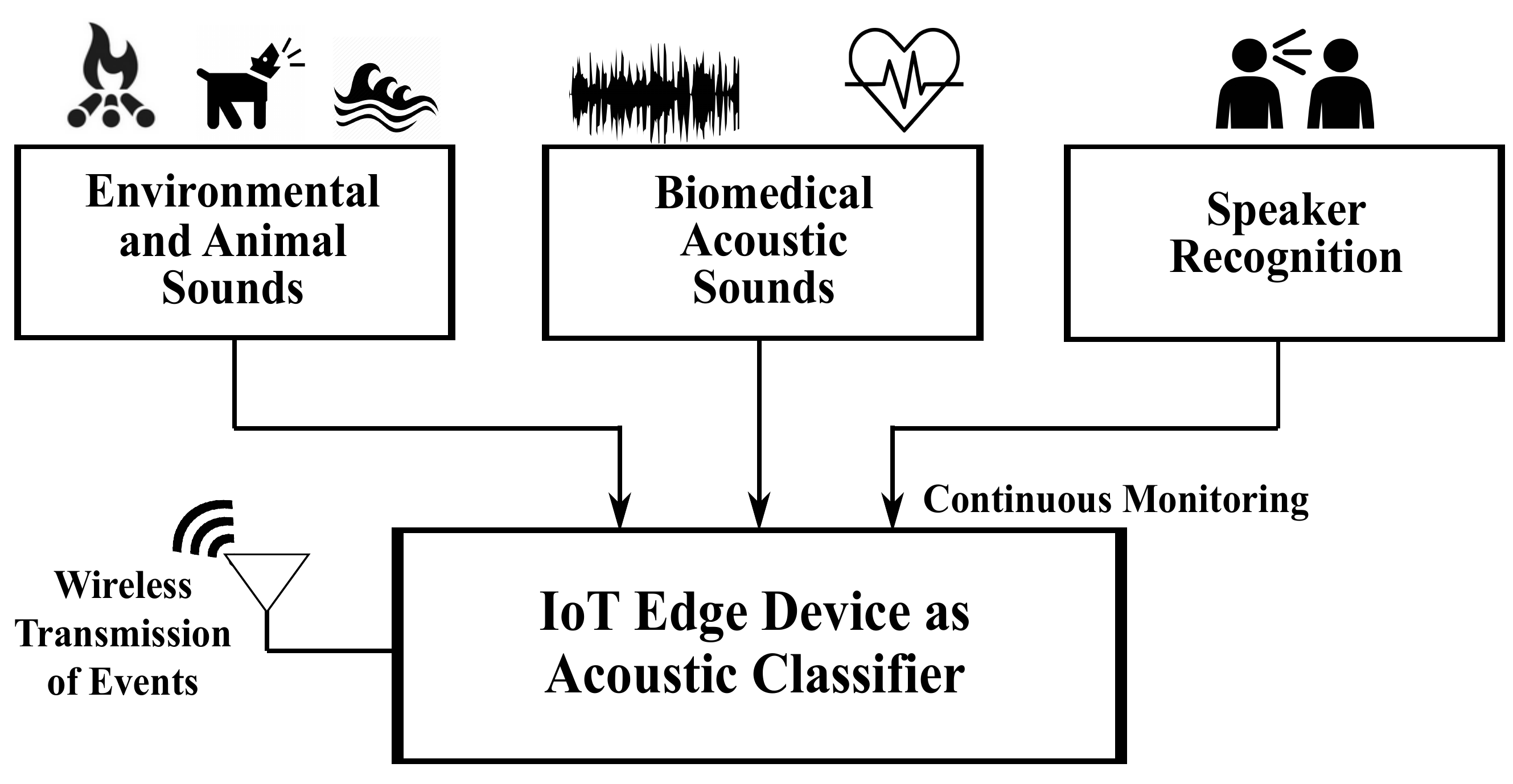}}
\caption{Smart IoT architecture using a classifier to reduce wireless transmission bandwidth }
\label{fig_motivation}
\end{figure}

\IEEEPARstart{I}{}nternet-of-Things like unattended ground sensors\cite{goodman1999detection} (UGS), intruder detection systems\cite{hoyt1994detection} \cite{valley1995method}, wild-life tracking ~\cite{zhong2017internet} or, structural health monitoring systems ~\cite{brown2000multi} generally operate in remote locations. They have to be active at all times to ensure that it can detect events of interest. In most cases, the events of interest are infrequent or rare. As a result, most of these IoT systems use an embedded pattern classifier to relax data storage and wireless transmission requirements~\cite{chakrabartty2007sub}. An example of such a system is illustrated in Fig.~\ref{fig_motivation}. Here the system wirelessly transmits alerts only when it detects signatures (acoustic or visual) about a target (for example a wild life species).

Based on this selective transmission, the IoT platform can conserve a significant battery power and hence prolong its operational life. However, the key challenge in designing such IoTs is that the integrated classifier needs to be robust and highly energy-efficient. While deep neural network (DNN) based classification systems can achieve very high accuracy\cite{hershey2017cnn}\cite{salamon2017deep}, there exist certain limitations when applying them to IoTs for rare-event detection. First, by the nature of the problem, the training data corresponding to the rare event is sparse and might not be suitable for DNNs. Even if it were possible to train a DNN for deployment, a compressed or a quantized variant of DNN, like the Binary Neural Networks (BNNs)\cite{BNN}, has to be used to optimize computational resources. Retraining BNNs on the IoT platform to account for data and hardware drifts is challenging due to quantization effects. Full-precision training is not possible due to limited computational resources. Also, if the parameters of the DNNs can be quantized, the input features cannot be significantly quantized without affecting the classification accuracy.  K-Nearest Neighbour (KNN) ~\cite{yang2020secure}\cite{quek2019iot} and Support vector machines (SVMs) \cite{tang2008svms}\cite{palaniappan2012abnormal}, on the other hand, have been shown to generalize well with sparse training data \cite{palaniappan2014comparative} . However, SVM is more robust to outliers, and the convexity of SVM training ensures any recalibration is interpretable and stable. There have been many instances where SVMs have performed well as an acoustic classifier \cite{temko2006classification} \cite{yue2018software}.  In literature, several approaches have been proposed to reduce the computational and memory footprint of SVMs~\cite{mahmoodi2011fpga}\cite{cutajar2013hardware}\cite{boujelben2018efficient}\cite{ramos2009svm}. However, in these platforms, computing features and classification are generally treated independently, both during training and inference.

In this paper, we present an in-filter computing framework that exploits the computing and nonlinear primitives in the feature extraction process to design 
ultra-light IoT acoustic classifiers. The approach is motivated by the fact that acoustic front-ends like the neuromorphic cochlea~\cite{xu2018fpga} can be designed to be highly computationally efficient using different degrees of linear and nonlinear transformations. Our goal is to systematically exploit and map these nonlinear transformations into the kernel functions used in SVMs, such that both classification and feature extraction are co-optimized for training and inference. This results in a template-based SVM~\cite{MWSCAS2019}\cite{GiniSVMMicro} architecture that has an ultra-low computational footprint for inference and training.  This feature not only relaxes communication bandwidth requirements on the IoT system but also allows recalibration (retraining) to account for statistical drifts. The main advantage of using template-based SVM is the ability of the framework to use arbitrary functions without any restriction on its properties, like positive-definite kernels for traditional SVM. This allows us to use hardware-friendly mapping or functions that need not be specified in a closed-form, such as using an ordinary differential equation (ODE). This property is beneficial especially for hardware implementation, where the inherent nonlinearity of the device can be used as a kernel rather than engineering a specific nonlinearity. As a proof-of-concept, we have applied the in-filter computing framework using an acoustic feature extractor based on Cascade of Asymmetric Resonators with Inner Hair Cells (CAR-IHC) \cite{xu2018fpga} \cite{lyon1998filter} \cite{lyon2017human}. The CAR-IHC model exhibits inherent nonlinearity and hence performs well as a kernel for classification. We believe that our
proposed framework has the following key advantages:
\begin{itemize}
  \item A template-based SVM architecture that allows an arbitrary function to be used as a kernel, unlike a conventional SVM that requires a positive-definite kernel.
  \item Combining the feature extraction and SVM kernel into one function makes the system ultra-light and computationally efficient.
  \item The memory footprint of the proposed system is user-defined and can be specified based on the IoT hardware constraints.
  \item A novel fast training algorithm with reduced training complexity in terms of memory and computational complexity.
  \item A system that can scale without affecting significant hardware changes due to the time-multiplexing approach allows the framework to deploy for more complex tasks. 
\end{itemize}
As proof of concept IoT implementation, we have implemented this inference framework on Xilinx Spartan 7 series FPGA~\cite{spartan7}, a low-cost and low-power FPGA. We have validated our architecture on various auditory datasets such as the environmental sound dataset \cite{piczak2015dataset} and speech-based dataset\cite{FSDD}.  

The rest of this paper is organized as follows. In Section II, a brief discussion of related work is provided, followed by section III, where we present the modified template-based SVM algorithm and explain the uniqueness of the formulation. In Section IV, we explain the novel training algorithm used for our framework. Section V provides the FPGA implementation details.
Section VI provides results obtained with an audio based dataset for detection and surveillance applications. Section VII concludes this paper and provides some useful applications, and discusses possible future work using this framework.

\begin{figure*}[ht]
    \centering
    \subfigure[]{\includegraphics[width=0.6\textwidth]{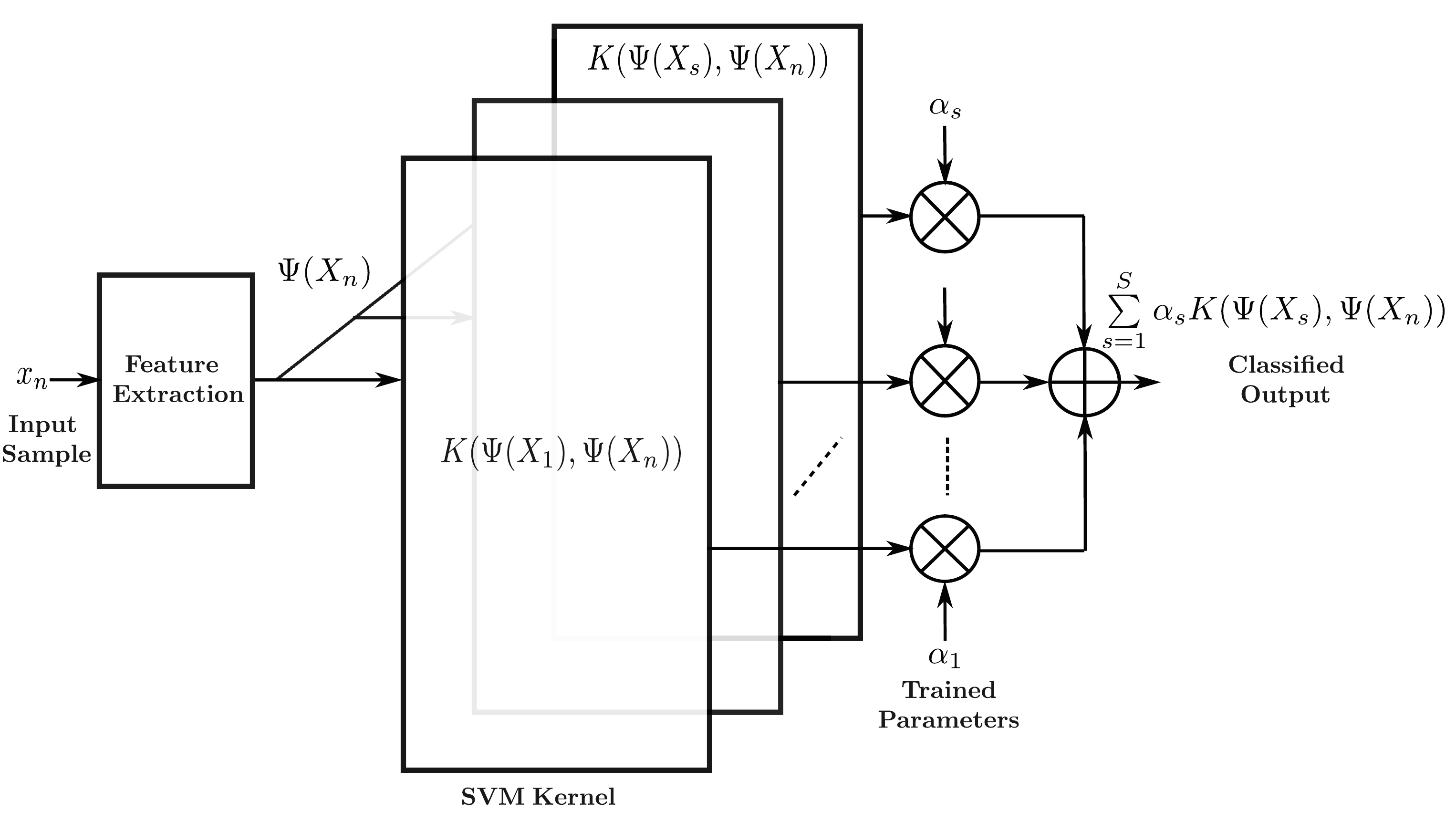}\label{fig:tradSVM}} 
    \subfigure[]{\includegraphics[width=0.38\textwidth]{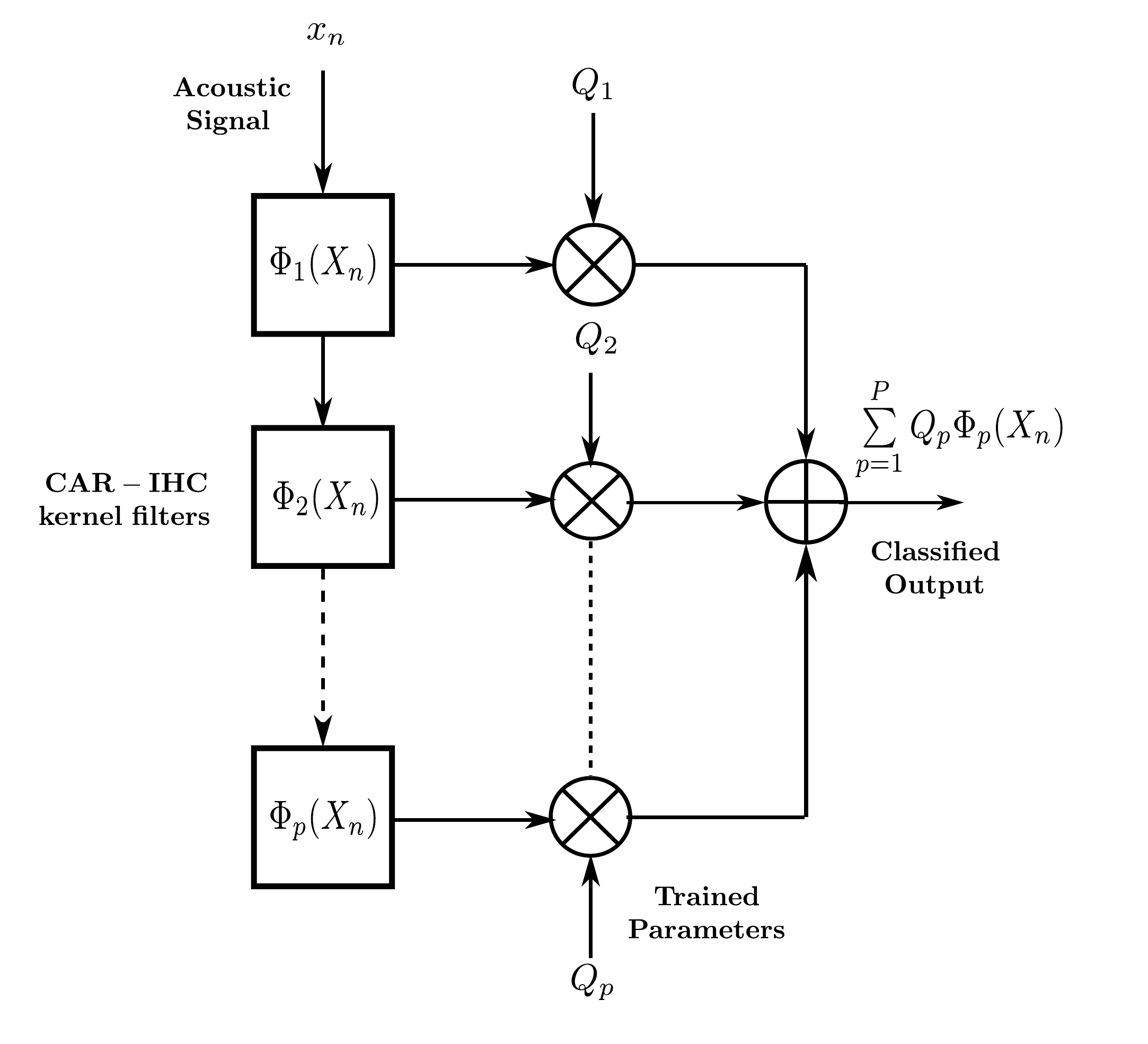}\label{fig:tempSVM}}
    \caption{Architecture of : (a) traditional SVM  based acoustic classifier (b) proposed in-filter SVM framework}
    
\end{figure*}

\section{Related Work}
Hardware implementations of SVM using FPGAs have been successfully achieved over the years with high accuracy and the least possible area and power. Binary classifications or even multi-class classifications using a Modified One-against-all (M-OAA) approach for SVMs have been implemented on FPGAs \cite{papadonikolakis2010novel} \cite{afifi2020fpga}. Since the kernel is one of the most important parts of the SVM algorithm, the kernel function consumes maximum resources in implementation. This is demonstrated in \cite{mahmoodi2011fpga} with linear and nonlinear SVM implementations on FPGA. The authors show that nonlinear kernel implementations use more resources than linear kernels. However, at the same time, there is a drop in accuracy by more than 10\% when using a linear kernel compared to a nonlinear kernel. The authors implement a kernel with parallel inputs enabling high operating frequency but at the cost of high resource utilization in terms of LUTs and DSPs. This shows that in order to get good classification, we require a nonlinear kernel, but at the same time, we need to achieve hardware efficiency for an ultra-light implementation.

Regarding acoustic feature extraction, acoustic signals require a certain amount of pre-processing to extract the salient features before it is used for classification. One such FPGA-based approach is detailed in \cite{cutajar2013hardware}. The authors use (Discrete Wavelet Transforms) DWTs for feature extraction from a given audio signal. This DWT feature extraction forms the input to a standard SVM having a Radial Basis Function (RBF) kernel, which is nonlinear. This  classification system is used for phoneme recognition using data from the TIMIT dataset. Due to hardware constraints and the complexity of the DWT algorithm, the authors chose to implement only the SVM classifier on the FPGA. The acoustic signals are pre-processed using a software implementation of the DWT algorithm and are provided as inputs to the SVM hardware. This implementation has the disadvantage of offline software feature extraction, making the hardware incapable of using unprocessed acoustic signals as inputs. At the same time, the SVM hardware implementation consumes a high number of FPGA resources in terms of LUTs and DSPs. Also, the weights and support vectors from the SVM training are stored in external ROMs. This makes the implementation impractical for a small IoT-based edge device.

Furthermore, time-series data need not always be a speech signal, and there may be cases where we may need to classify non-auditory time-series signals. In \cite{boujelben2018efficient}, authors use Mel-frequency cepstral coefficients (MFCC) technique to extract salient features from pulmonary sounds to detect wheezing using standard SVM classification. Here, MFCC, as well as SVM, was implemented on FPGA. This implementation provided an end-to-end solution on hardware that could classify between a normal and an abnormal pulmonary sound. In this implementation, MFCC itself is a resource-heavy algorithm, and additional hardware is required for the SVM classifier to be implemented. Also, ROMs store support vectors and weights along with additional registers to store MFCC coefficients. The MFCC coefficient calculations, which are being done on hardware, also contribute to high DSP usage. The authors have demonstrated their hardware capability using only a 6 kHz input sampling frequency, making the hardware limited in terms of the flexibility of signals that it can process. Hence, such a system cannot be used in an IoT edge device due to the high resource utilization and rigidity.

Another representative example of an IoT for acoustic classification is a speaker identification system used in security systems. One such system was realized on FPGAs in \cite{ramos2009svm}. Similar to the implementation in \cite{boujelben2018efficient}, the authors implemented an SVM classifier with MFCC as the feature extractor on hardware. The input data was sampled at 8 kHz, making it resource-efficient, but at the same time, it was less flexible in terms of processing signals of higher sampling frequency. External SRAM was used to store the MFCC coefficients and training parameters. Despite having a slight improvement in terms of hardware efficiency compared to the previous implementation, this implementation lacked flexibility and still had a significant amount of resource usage, given the hardware constraints applicable for an IoT device.

Our framework addresses all the shortcomings of prior works by having a neuromorphic cochlea-based CAR-IHC kernel integrated inside a template-based SVM system. This kernel exhibits nonlinearity for better classification and, at the same time, inherently provides a robust feature extracting capability in order to get a good classification. This kernel has multiple tunable parameters which can be adjusted to get the best feature extraction depending on the application. The template-based SVM provides the flexibility of choosing the right number of templates as support vectors, which can be tuned as per the application. This avoids the additional requirement for the  storage of support vectors. Flexibility, scalability, low resource usage, and low power make this framework ultra-light and ideal for IoT deployment for many applications.

\section{Template-based SVM and In-filter Computing Formulation}
Rooted in statistical learning theory, an SVM minimizes the structural risk by maximizing a classification margin over a set of training samples\cite{cortes1995support}. 
In the case of acoustic classification where the input is a time-series signal, one can define a data vector (for training and inference) at a time-instant $n$ as $X_n \in \mathbb{R}^W$ constructed using a window $X_n = \{x_n, x_{n-1},...,x_{n-W+1} \}$ of previous $W$ samples of the signal $x_n$. An SVM based binary classifier produces a
decision label $y_n \in \{+1, -1\}$ corresponding to the data vector $X_n$ according to
\begin{equation}
y_n = \sgn(f(X_n)).    \label{eq:1} 
\end{equation}
where $f:\mathbb{R}^W \rightarrow \mathbb{R}$ is given by
\begin{equation}
f(X_n) = \sum\limits_{s=1}^S \alpha_{s}y_{s}K(X_{s},X_n) + b.  \label{eq:2}      
\end{equation}
Here $X_s \in \mathbb{R}^W$ is a subset of the training vector called support vectors with their a-priori known decision labels $y_s \in \{+1,-1\}$.
$K:\mathbb{R}^W \times \mathbb{R}^W \rightarrow \mathbb{R}$ is a positive-definite kernel function that is also chosen a-priori and plays
an important role in implementing nonlinear decision functions.  
$\alpha_{s} \in \mathbb{R}$ and $b$ are training parameters, corresponding to the support vector $x_s$ and is determined by solving a 
standard quadratic program based training procedure\cite{cortes1995support}.
Note that the memory requirements to implement an SVM inference engine in hardware is proportional to the number of support vectors $S$,
and hence in literature numerous techniques exist to reduce $S$ using heuristic methods~\cite{habib2009support} \cite{wang2004heuristic}.

For conventional SVM-based acoustic classifiers\cite{guo2003content}, as shown in Fig.\ref{fig:tradSVM}, the raw input signal is pre-processed by a feature extraction module or function
 $\Psi:\mathbb{R}^W \rightarrow \mathbb{R}^D$ before providing as an input to the SVM kernel. $D$ is the feature dimension. The eq.\eqref{eq:2} can be re-expressed as
 \begin{equation}
    f(X_n) = \sum\limits_{s=1}^S \alpha_{s}y_{s}K(\Psi(X_{s}),\Psi(X_n)) + b.  \label{eq:2a}      
\end{equation}
In literature, the kernel function $K(.,.)$ and the feature extraction function $\Psi(.)$ are typically chosen independently. As a result, the memory footprint $S$ of the SVM is determined by the complexity of the problem and the discriminative power of feature extraction. Note that a typical acoustic 
feature extraction function $\Psi(.)$, itself comprises several nonlinear transformations that could directly be used as SVM kernels. However, for the
SVM formulation to be valid, the nonlinear transformations must be mapped to a positive-definite kernel. In~\cite{MWSCAS2019} we reported a mechanism to design SVMs using arbitrary template functions within
a fixed memory footprint. The approach expressed the kernel in eq.\eqref{eq:2}, as an outer-product over $P$ template functions $\Phi_p:\mathbb{R}^W \rightarrow \mathbb{R}$ as $K(X_s,X_n)= \sum \limits_{p=1}^P \Phi_p(X_s) \Phi_p(X_n),$. The template functions $\Phi_p(\cdot), p = 1,..,P$ then could represent $P$
feature extraction modules. Following the derivations in \cite{MWSCAS2019}, the SVM function $f(.)$ can be rewritten as:

\begin{eqnarray}
    f(X_n) = \sum \limits_{p=1}^P Q_p \Phi_p(X_n) + b. \label{eq:5}   
\end{eqnarray}
Here, $Q_p=\sum \limits_{s=1}^S \alpha_{s}y_{s}\Phi_p(X_s)$ can be viewed as a consolidated training parameter that can be estimated using a reduced-complexity
training procedure described in section~\ref{Training}. Note that the memory footprint of the reformulated template-SVM is determined by the number of template
functions $P$, and each of the template functions $\Phi_p(\cdot)$ could be chosen arbitrarily. Here, $\Phi_p(\cdot)$ can be any function that can be used to express the input features in order to make a classification. This makes this framework flexible to implement various functions for classification. For example, in Fig.\ref{fig:tempSVM}, we illustrate how a cascade of filters a CAR-IHC feature extraction module could be used to implement $\Phi_p(\cdot)$, and described in the following section.

\begin{figure}[ht]
    \centering{\includegraphics[page=1,scale=0.4,trim=0 0 0 0,clip]{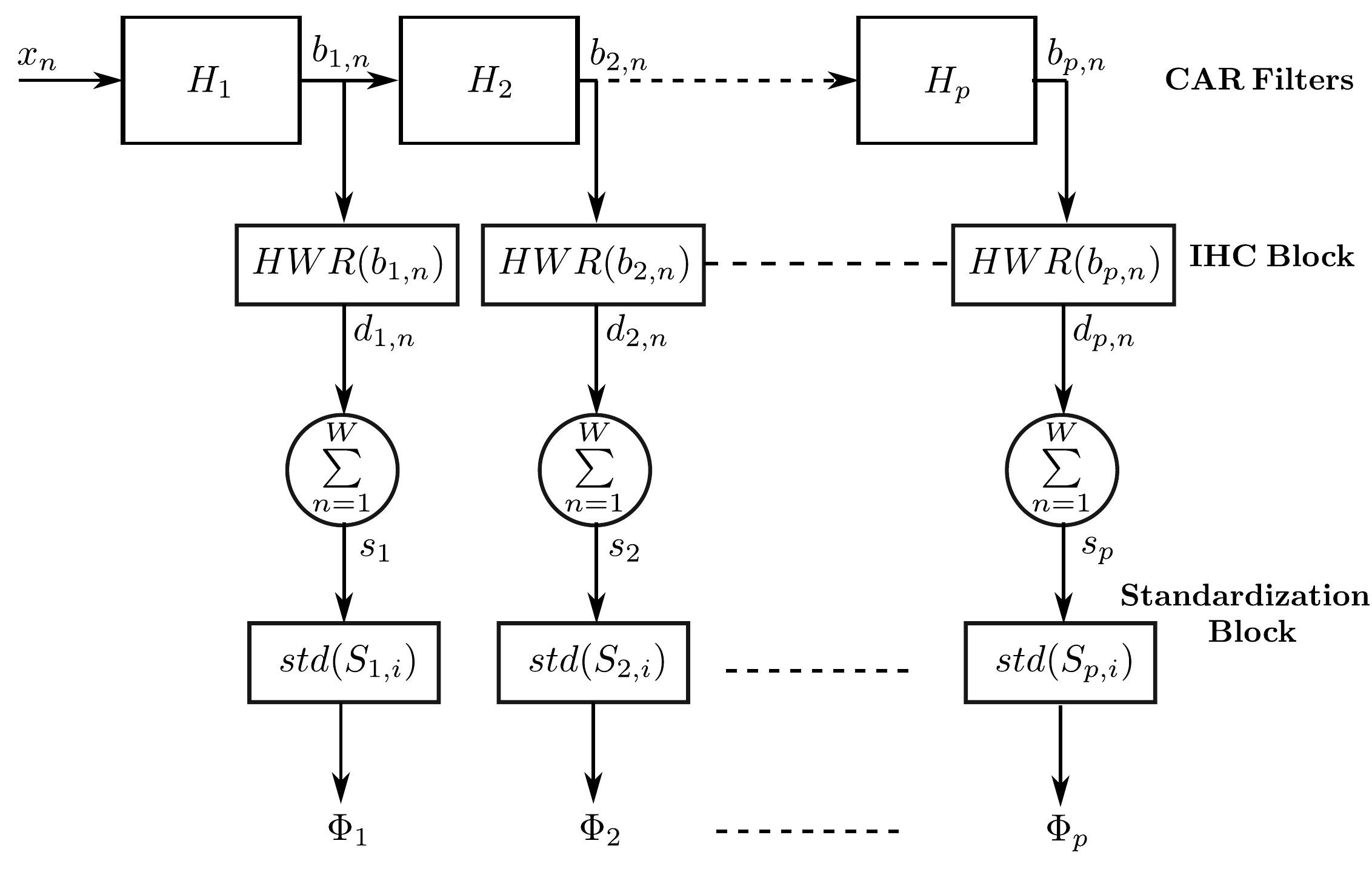}}
    \caption{Neuromorphic cochlea-based CAR-IHC model }
    \label{Kernel}
\end{figure}

\subsection{CAR-IHC model as SVM Kernel}
The biological cochlea is a nonlinear and causal system. This nonlinearity makes it ideal to use a cochlea model as an SVM kernel since it would give robust classification in higher dimensional space \cite{girolami2002mercer}. One such auditory filter model is the  Cascade of Asymmetric Resonators with Fast-Acting Compression (CARFAC) model, which is a digital version of the cascade of pole-zero filter \cite{xu2018fpga}\cite{lyon2017human} \cite{thakur2014fpga}. It consists of CAR block, which mimics Basilar Membrane (BM) functionality, IHC, Ganglion cells, and  Outer Hair Cell (OHC). We use the CAR and IHC modules of this model for our kernel.

Given $X_n$ with $W$ samples and each sampled input data as $x_{n}$ described previously. The system receives an audio sample at each sampling clock which is fed to the first CAR block $H_1$.
There are $P$ CAR blocks arranged in a cascaded manner as shown in Fig.\ref{Kernel}.
The eq.\eqref{eq:8a} denotes the two pole two zero filter which mimics BM implemented as a CAR filter.

\begin{equation}
     H_p = H(z) = g_p\left[\frac{z^{2}+(-2a_{0,p}+k_pc_{0,p})r_pz+r_p^{2}}{z^{2}-2a_{0,p}r_pz+r_p^{2}}\right]. \label{eq:8a}
\end{equation}
$a_{0,p},c_{0,p},k_p$ are the resonator filter coefficients for each filter, $r_p$ is the pole-zero radius in the z-plane, and $g_p$ is the DC gain factor. The CAR block transfer function $H(z)$ is derived in detail in Appendix. 

For the first CAR filter, the input is $x_n$, and due to the cascade nature of the CAR filters, the output of one filter becomes the input of the next stage filter.
The output of each CAR filter is denoted by $b_{p,n}$ as shown in Fig.\ref{Kernel}, which forms the input to the IHC blocks in parallel.
We use a simplified model of the IHC implemented using half wave rectifier (HWR), $HWR(\cdot) \in \mathbb{R}$, as per eq.\eqref{eq:11}. 
\begin{equation}
    HWR(q) = max(0,q). \label{eq:11}
\end{equation}

From Fig.\ref{Kernel}, $q = b_{p,n}$ in eq.\eqref{eq:11}, which gives,
\begin{equation}
    d_{p,n} = HWR(b_{p,n}). \label{eq:9a}
\end{equation}
The IHC generates output as per eq.\eqref{eq:9a}. The IHC output is summed over $W$ samples, and this forms the input for the standardization (std) blocks in parallel.  

\begin{equation}
    s_p = \sum\limits_{n=1}^{W}d_{p,n}. \label{eq:12a}
\end{equation}
Here, $s_p \in \mathbb{R}$.
\begin{equation}
    std(S_{p,i})=\frac{S_{p,i}-\mu_p}{\sigma_p}.  \label{eq:12} 
\end{equation}
where $\{ s_p \in S_{p,i} | 1 \leq i \leq N \}$ with $N$ as the training samples, $\mu_p = mean(S_{p,1}, S_{p,2}, .., S_{p,N})$ and  \\ $\sigma_p$ =$\sqrt{\frac{1}{N-1}\sum\limits_{i=1}^N(S_{p,i}-\mu_p)^2}$

\begin{equation}
   \Phi_p = std(S_{p,i}).\label{eq:6}
\end{equation}
Here, $\Phi_p  \in  \mathbb{R} $.
The summation over $W$ samples of the output of IHC is taken as per eq.\eqref{eq:12a} for each filter. Then standardization technique, commonly used in neural network optimizations ~\cite{shanker1996effect}, is applied across $N$ training input samples as per eq.\eqref{eq:12}. Note that $\mu_p$ and $\sigma_p$ are calculated only during training, and these vectors are passed as learned parameters to the inference engine. 
Therefore, an input signal vector $X_n$ sampled at a sampling frequency $f_{s}$ generates $W$ samples with each sample denoted as $x_{n}$. It is then processed by a cascade parallel arrangement of neuromorphic cochlea-based CAR-IHC filters to estimate the kernel function $\Phi_p$ with ${p}$ as the filter stage out of $P$ filters as per \eqref{eq:6}. 
The output is a $P\times 1$ kernel vector, as shown in Fig.\ref{Kernel}. The classification output is produced using eq.\eqref{eq:5} employing this kernel vector, the output weight vector $Q \in \mathbb{R}^P$, and the bias $b \in \mathbb{R}$ obtained after the training process. 

\section{Template-SVM Training} \label{Training}

A conventional SVM training involves solving a quadratic optimization problem over a set of training data $(X_m,y_m), m=1,..,M$, of
size $M$ \cite{cortes1995support}. The optimization can be expressed as:
\begin{eqnarray}
    \underset{\alpha_m}{min} \; \frac{1}{2}\sum\limits_{m=1}^M\sum\limits_{n=1}^M \alpha_m \alpha_n y_m y_n K(X_n,X_m) - \sum\limits_{m=1}^M \alpha_m. \label{eq:13} \\
    \sum\limits_{m=1}^M \alpha_m y_m = 0. \qquad \qquad \qquad \quad \label{eq:15} \\
    0 \leq \alpha_m \leq C. \qquad \qquad \; \; \qquad \quad\label{eq:16}
\end{eqnarray}
Here, $C$ is a hyper-parameter that is chosen through cross-validation and $n=1,..,M$. Due to the quadratic nature of the optimization problem, the worst-case complexity of SVM training scales as $\mathcal{O}(M^2)$. In practice, the training complexity scales as $\mathcal{O}(MS)$, where $S$ is the number of support vectors. However, the number of support vectors is unknown, so any SVM formulation has to accommodate the worst-case scenario.

In the template-based SVM, the kernel is expressed as an outer-product over a set of $P$ templates as $K(X_n,X_m)= \sum \limits_{p=1}^P \Phi_p(X_n) \Phi_p(X_m)$. Substituting in equation~\eqref{eq:13}, the template-SVM training reduces to a lower complexity quadratic optimization problem as:  
\begin{eqnarray}
    \underset{Q_p,\alpha_m}{min} \; \frac{1}{2}\sum\limits_{p=1}^P Q_p^2 - \sum\limits_{m=1}^M \alpha_m. \label{eq:17} \\
    s.t \;\; Q_p = \sum\limits_{j=1}^M \alpha_j y_j \Phi_p(X_j).\; \label{eq:18} \\
    \sum\limits_{m=1}^M \alpha_m y_m = 0. \qquad \quad \label{eq:19} \\
    0 \leq \alpha_m \leq C. \qquad \quad \; \; \label{eq:20}
\end{eqnarray}
Here, $j$ is iterated over $M$ training samples. Equations eq.\eqref{eq:18}, \eqref{eq:19}, \eqref{eq:20} are the constraints imposed on eq.\eqref{eq:17}.
The equation eq.\eqref{eq:17} shows that the optimization complexity has been reduced to $\mathcal{O}(P+M)$ with $P$ additional constraints that can be controlled based on the number of templates required for the application. This reduced complexity enables us to use this training algorithm to implement IoT devices, making them adaptive and deployable in dynamic environments. Thus, our framework is capable of online training. For the in-filter computing, the features or templates are computed as an input stream. Training the template SVM entails solving a simplified constrained quadratic problem. Thus, the architecture can be trained in an online manner. However, in a traditional SVM, the features would first need to be accumulated and fed to a kernel module. Note that there are several ways to efficiently solve the constrained optimization in eq.\eqref{eq:17}, including both batch and online variants. In the accompanying software for template-based SVM\cite{GiniSVMMicro}, we have used a growth-transform-based approach~\cite{gangopadhyay2017extended} to solve eq.\eqref{eq:17}.

\begin{figure}[ht]
\centerline{\includegraphics[page=1,scale=0.5,trim=0 0 1 0,clip]{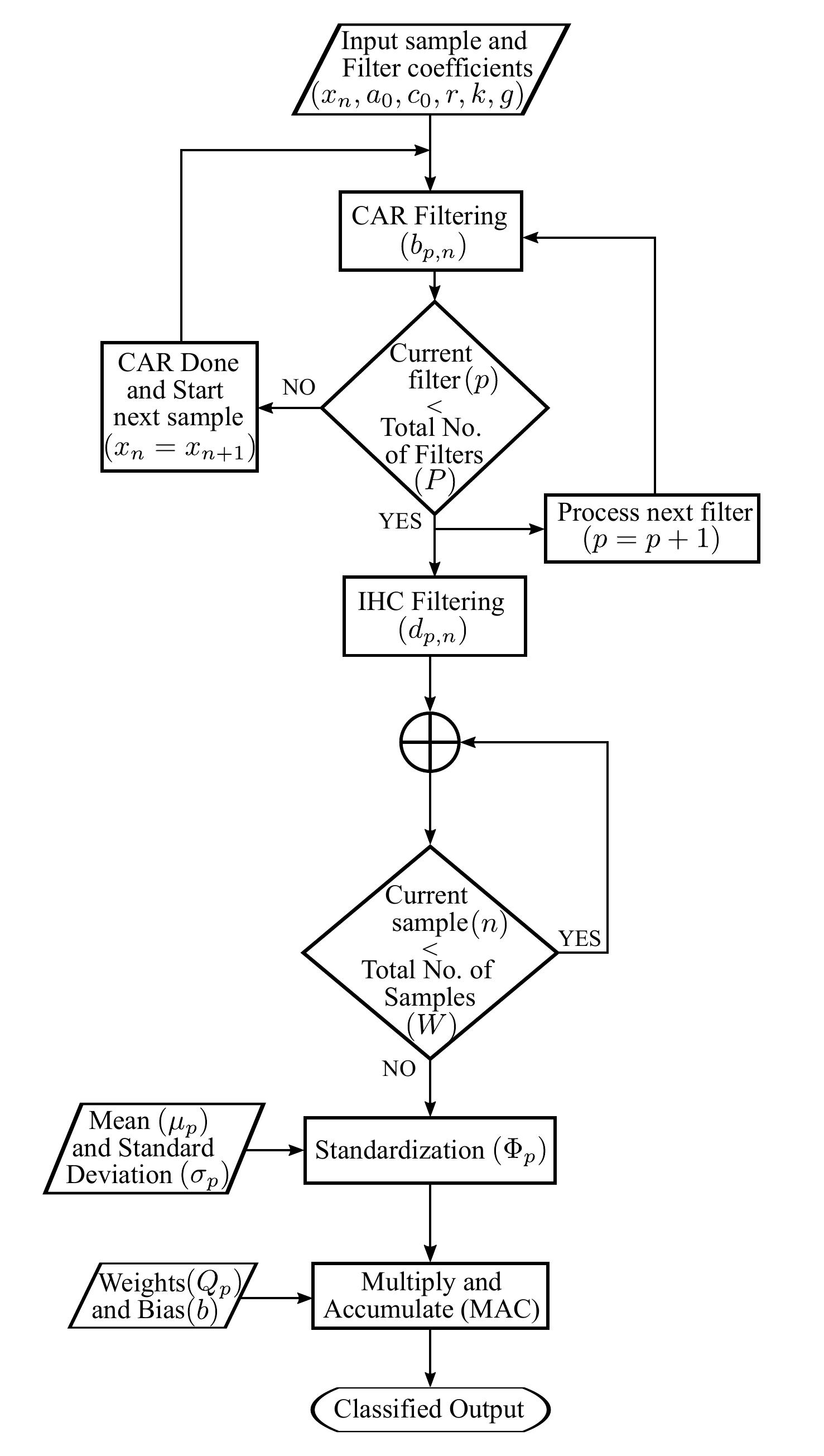}}
\caption{High level hardware execution flow diagram of the framework.}
\label{Flow_Diag}
\end{figure}

\section{FPGA Implementation of our SVM classifier}\label{FPGA_Impl}
We demonstrate the efficiency of our in-filter computing architecture by implementing it on FPGA. This system is configurable as the in-filter parameters, and weight vectors are tunable based on the application. The weight vector and biases are trained offline, as mentioned in the previous sections. 
\IEEEPARstart{} Initially, we simulated this framework in floating-point in MATLAB software tool. In order to estimate the appropriate FPGA implementation, we simulated the model in fixed-point code. The CAR-IHC kernel is implemented in a 12-bit fixed-point code. The trained weights ($Q_p)$ and bias ($b$) are stored as 8-bit, and the mean ($\mu_p$) and standard deviation ($\sigma_p)$ are stored as 12-bits. In our experiments, we analyzed that using 12-bits for inputs, filter coefficients and standardization parameters (mean and standard deviation) with 8-bits for weights and bias resulted in minimal accuracy degradation and reduced hardware resource utilization. We use pipelining to speed up the kernel execution in FPGA. The CAR-IHC kernel filters are executed using the time-multiplexed technique where each filter uses the same hardware for generating output which makes the design small in area.
\begin{table}[!ht]
\centering
\caption{FPGA implementation summary.} \label{FPGA_sum}
\begin{tabular}{|c||c|}
\hline
\multicolumn{2}{|c|}{\textbf{\begin{tabular}[c]{@{}c@{}}FPGA Implementation Summary \\ Spartan-7 xc7s6cpga196\end{tabular}}} \\ \hline \hline
Clock Frequency                      & 25 MHz              \\ \hline
Audio Sampling Frequency             & 16 kHz              \\ \hline
Number of Filters                    & 30                  \\ \hline
Dynamic Power                                & 8 mW            \\ \hline
DSPs                           & 4 (Total 10) \\ \hline
LUTs                           & 1517 (Total 3750)              \\ \hline
FFs                            & 2864 (Total 7500)  \\ \hline

\end{tabular}

\end{table}

\begin{table*}[ht]
\centering
\caption{Comparison of architecture and resource utilization of related work.\label{RLUT}}
 \begin{threeparttable}[t]
\begin{tabular}{|l||c|c|c|c|c|}
\hline
\textbf{\begin{tabular}[c]{@{}c@{}}Related Work\end{tabular}} &
  \textbf{\begin{tabular}[c]{@{}c@{}}Mahmoodi, et al.\\ \cite{mahmoodi2011fpga}\end{tabular}} &
  \textbf{\begin{tabular}[c]{@{}c@{}}Cutajar, et al.\\ \cite{cutajar2013hardware}\end{tabular}} &
  \textbf{\begin{tabular}[c]{@{}c@{}}Boujelben, et. al.\\ \cite{boujelben2018efficient}\end{tabular}} &
  \textbf{\begin{tabular}[c]{@{}c@{}}Ramos-Lara et al.\\ \cite{ramos2009svm}\end{tabular}} &
  \textbf{This work} \\ \hline  \hline
\textbf{FPGA}                         & Virtex4-xc4vsx35          & Virtex-II XC2V3000       & Artix-7 XC7A100T  & Spartan 3 XCS2000           & Spartan 7 xc7s6cpga196                                                                   \\ \hline
\textbf{Operating Frequency}          & 151.286 MHz              & 42.012 MHz               & 101.74 MHz                & 50 MHz                      & 25 MHz                                                                         \\ \hline
\textbf{Input Sampling Frequency}     & NA\tnote{1}             & 16 kHz                    & 6 kHz                 & 8 kHz                       & 16 kHz                                                                         \\ \hline
\textbf{Flip Flop}                    & 11589                   & 1576                     & 17074                  & 5351                        &   2864                                                                         \\ \hline
\textbf{LUTs}                 & 9141                    & 11943                    & 16563          & 6785                        &1517                                                                           \\ \hline
\textbf{RAM (18 Kb)}                          & 99                       & NA\tnote{1}               & 4                                                 & NA\tnote{1}                          & 0                                                                              \\ \hline
\textbf{DSP}                          & 81                        & 64                      & 87                                               & 21                          & 4                                                                             \\ \hline
\textbf{Kernel type}                  & Gaussian                  &RBF                      & Linear          & RBF                         & CAR-IHC                                                                        \\ \hline
\textbf{Feature extraction}           & NA\tnote{1}                & DWT (Offline)        & MFCC (On-Chip)                         & MFCC (On-chip)              & CAR-IHC (On-chip) \\ \hline
\end{tabular}
\begin{tablenotes}
     \item[1] These works did not report this entity for their designs.
\end{tablenotes}
\end{threeparttable}%

\end{table*}

\begin{figure}[ht]
\centerline{\includegraphics[page=1,scale=0.47,trim=0 0 0 0,clip]{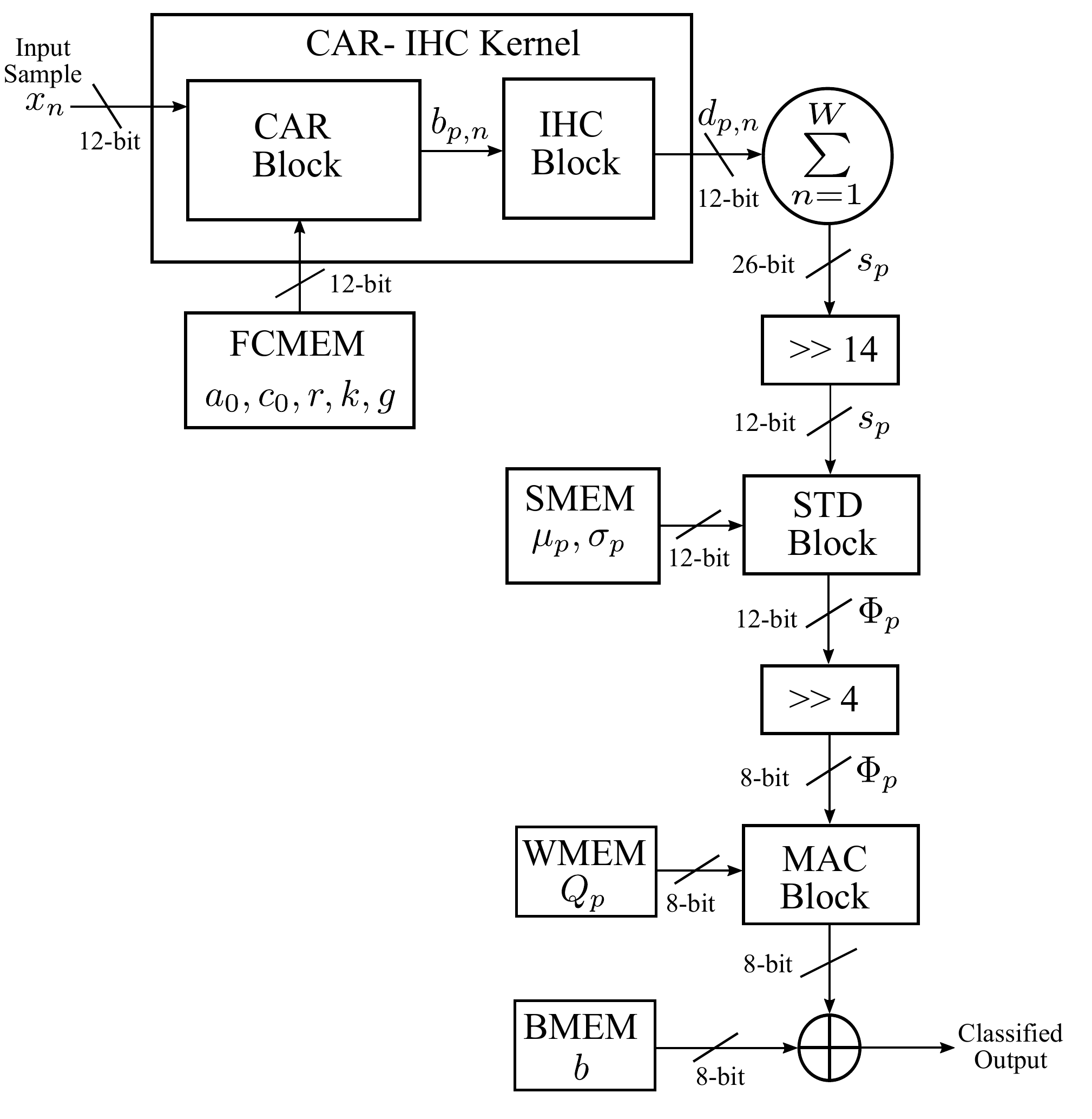}}
\caption{ Block diagram for FPGA implementation }
\label{FPGA}
\end{figure}

\begin{figure}[ht]
\centerline{\includegraphics[page=1,scale=0.4,trim=0 0 1 0,clip]{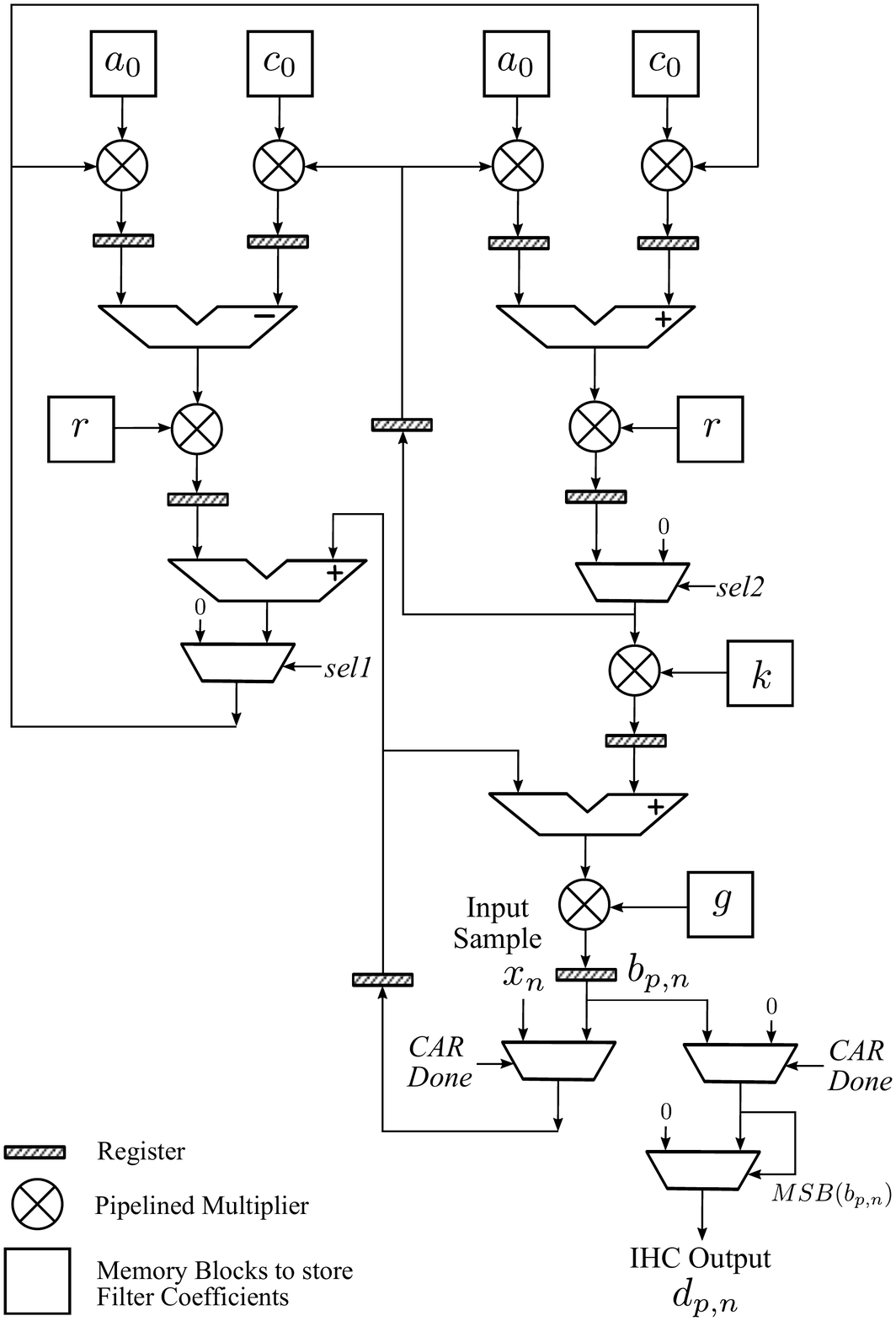}}
\caption{ CAR-IHC kernel hardware micro-architecture implementation. The filter coefficients ($a_0,c_0,r,k,g$) stored in block RAM and the input sampled at $f_s$ are used as inputs to this block. This block is used in pipelined manner using select lines to enable the FSM for the design.}
\label{FPGA_CARIHC}
\end{figure}

Fig.\ref{Flow_Diag} shows the hardware execution flow, and Fig.\ref{FPGA} shows the FPGA block architecture of the system. The input sample ($x_n$) sampled at $f_s$ is provided as input to CAR block and filter coefficients ($a_0, c_0, r, k, g$), stored in the FCMEM memory block. The CAR block performs filtering as per eq.\eqref{eq:8a} followed by the IHC block, which performs half wave rectification as per eq.\eqref{eq:11}. The detailed implementation of the CAR-IHC block is shown in Fig.\ref{FPGA_CARIHC}. The time-multiplexed processing of each cascaded filter is determined by the $CAR$ $Done$ select signal. If this signal is set, then the next sample ($x_{n+1}$) enters the CAR block. This signal is set only when the processing of all the filters is done. Hence, the same CAR block is used to process $P$ filters using the stored filter coefficients for each filter ($p$). The $sel1$ and $sel2$ select lines control this filter coefficient flow. So the input to output delay is directly proportional to the number of cascaded blocks, $P$ in this case, i.e., the number of filters used. The multipliers have been designed to operate in a pipelined manner, where multiplication of the coefficients will take multiple clock cycles for producing the output by reusing the hardware. The half wave rectification operation in the IHC block depends on the Most Significant Bit ($MSB$) select signal, determining the sign bit of CAR output ($b_{p,n}$). This results in discarding the values below zero to produce the rectified output. 

The output of IHC block ($d_{p,n}$) is summed over the entire window of the input data, i.e., summation of $W$ input samples across $p$ filters. Standardization of the output ($s_p$) of summation is performed based on eq.\eqref{eq:12} in the STD block. The size of $s_p$ is 26-bit due to 16000 ($f_s$)  additions, and the standardization parameters, i.e., mean ($\mu_p$) and standard deviation ($\sigma_p$), are 12-bits, which are fetched from the SMEM memory block. We quantize $s_p$ to 12-bit and further perform another level of quantization to 8-bit after the standardization operation. Heuristically, we found that using these multiple quantization levels achieves the lowest possible hardware resource utilization without impacting the classification accuracy. The output of standardization block ($\Phi_p$) is now used to perform multiply-accumulate (MAC) operation with the learned weights ($Q_p$), which are also 8-bits, stored in the WMEM memory block. Finally, the classified output is obtained after the bias ($b$), stored in the BMEM memory block, is added to the output of the MAC block, as per eq.\eqref{eq:5}.

Our system uses a 25 MHz system clock, and 16-bit input sampled at a 16 kHz audio sampling rate. We have used 30 filters in all the reported results here. However, it is parameterizable and can be changed based on the application requirements. Every input data sample  takes about 300 clock cycles, i.e., 12$\mu$s, to be executed through the pipeline. We have an audio input sample being given to the system every 1562 clock cycles, i.e., 62.5$\mu$s. We have a buffer of around 1200 clock cycles, i.e., about 48$\mu$s, before the arrival of next sample, where the, system is idle. This shows that we can increase the sampling frequency to 80 kHz, i.e., sample at every 312 clock cycles without impacting the hardware architecture. On the other hand, for a sample of 16 kHz frequency, we can also increase the number of filters to up to 120 to use up the extra 48$\mu$s. The number of clock cycles required to execute a single audio sample increases linearly with the number of filters. There is an increase of about 2.5 \% of area in overall hardware and 0.23 mW increase in power for every addition of filter. The increase in the number of filters may be required for complex auditory tasks. We can also reduce the operating frequency to as low as 400 kHz to reduce the power consumption of the system to below 1 mW. We can reduce it further to a few kHz if we reduce the input sampling frequency to a few Hz in other time-series data such as EEG/ECG signals. Hence, this shows that our system is highly flexible and scalable to suit any application in the time-series domain.

\begin{table*}[ht]
\centering

\caption{ESC-10 dataset classification accuracy results in percent. The fixed point code consists of 16-bit inputs and 8-bit weights and biases. Number of filters for our work is fixed at 30. \label{ESC10}}

\begin{tabular}{|c||c|c|c|c|c|c|c|}
\hline
\multirow{3}{*}{\textbf{Classes (Train/Test)}} & \multicolumn{3}{c|}{\textbf{Traditional SVM}}                                                                                          & \multicolumn{4}{c|}{\textbf{In-filter SVM (This Work) }}                                                                              \\ \cline{2-8} 
                                               & \multirow{2}{*}{\textbf{\begin{tabular}[c]{@{}c@{}}Support \\ Vectors\end{tabular}}} & \multicolumn{2}{c|}{\textbf{Floating pt.}} &
                                               \multicolumn{2}{c|}{\textbf{Floating pt.}} & \multicolumn{2}{c|}{\textbf{Fixed pt.}} \\ \cline{3-4} \cline{5-8} 
                                               &                                                                                      & \textbf{Train}       & \textbf{Test}                                          & \textbf{Train}       & \textbf{Test}       & \textbf{Train}      & \textbf{Test}     \\ \hline \hline
\textbf{Dog (129/33)}                          & 51                                                                                   & 88                    & 87                                                 & 89                   & 90                  & 89                 & 87                \\ \hline
\textbf{Rain (119/40)}                         & 60                                                                                  & 89                   & 87                                                   & 89                   & 87                  & 82                  & 82               \\ \hline
\textbf{Sea\_Waves (200/50)}                   & 113                                                                                   & 86                   & 82                                                  & 84                   & 78                  & 80                  & 74                \\ \hline
\textbf{Crying Baby (144/49)}                  & 58                                                                                   & 91                   & 83                                                   & 91                  &  87                  & 93                  &  85                \\ \hline
\textbf{Clock Tick (114/50)}                   &  86                                                                                  & 91                   & 86                                                  & 92                   & 88                  & 92                  & 85              \\ \hline
\textbf{Person Sneeze (101/44)}                & 43                                                                                  & 83                  & 77                                               & 89                   & 82                  & 82                  & 80               \\ \hline
\textbf{Helicopter (197/50)}                   & 48                                                                                   & 94                  & 88                                                   & 96                   & 90                  & 95                 & 85               \\ \hline
\textbf{Chainsaw (99/34)}                      & 39                                                                                   & 92                  &  85                                                  & 93                   & 85                  & 93                  & 82                \\ \hline
\textbf{Rooster (124/54)}                      & 36                                                                                   & 92                   & 94                                                  & 93                   & 96                 & 93                  & 96                \\ \hline
\textbf{Fire Crackling (152/66)}               & 46                                                                                   & 90                 & 89                                                 & 90                  & 87                 & 89                  & 86                \\ \hline
\end{tabular}

\end{table*}

\begin{table*}[ht]
\centering

\caption{FSDD classification accuracy results in percent. The fixed point code consists of 16-bit inputs and 8-bit weights and biases. Number of filters for our work is fixed at 30.\label{FSDD}}

\begin{tabular}{|c||c|c|c|c|c|c|c|c|}
\hline
\multirow{3}{*}{\textbf{Classes (Train/Test)}} & \multicolumn{3}{c|}{\textbf{Traditional SVM}}                                                                                          & \multicolumn{4}{c|}{\textbf{In-filter SVM (This Work) }}                                                                              \\ \cline{2-8} 
                                               & \multirow{2}{*}{\textbf{\begin{tabular}[c]{@{}c@{}}Support \\ Vectors\end{tabular}}} & \multicolumn{2}{c|}{\textbf{Floating pt.}} &
                                               \multicolumn{2}{c|}{\textbf{Floating pt.}} & \multicolumn{2}{c|}{\textbf{Fixed pt.}} \\ \cline{3-4} \cline{5-8} 
                                               &                                                                                      & \textbf{Train}       & \textbf{Test}                                          & \textbf{Train}       & \textbf{Test}       & \textbf{Train}      & \textbf{Test}     \\ \hline \hline
\textbf{Theo (761/254)}                        & 247                                                                                  & 92                   & 91                                               & 93                    & 91                  & 89                & 88                \\ \hline
\textbf{Nicolas (889/297)}                     & 197                                                                                  & 98                   & 98                                                  & 98                   & 97                 & 97                  & 94                \\ \hline
\textbf{Ywewelver (749/250)}                     & 196                                                                                  & 92                  & 90                                                 & 94                   & 91                  & 89                  & 88               \\ \hline
\textbf{Jackson (796/200)}                     & 35                                                                                  & 99                   & 99                                                   & 99                   & 99                  & 99                  & 98               \\ \hline
\end{tabular}

\end{table*}

We use Xilinx Spartan series part xc7s6cpga196, a low-power FPGA manufactured on a 28 nm technology node.The Spartan series FPGA caters to edge computing and IoT platform systems, as the area footprint and power envelope are low. The dynamic power consumption for our system on FPGA is 8 mW. The logic design i.e., LUTs and registers, consumes around 2 mW of power, whereas the control signals take up 4 mW. DSPs consume 1 mw of power, and the clocks take up the remaining 1 mw of power. The weights and bias generated from the training procedure are quantized to 8-bit, and the CAR-IHC model uses 16-bit input samples to generate 16-bit output. This 16-bit kernel output data on accumulation over 16k samples increases to 30-bit, which is then reduced to 8-bit after a standardization and quantization operation. For this design, 16 kHz sampling rate with 30 filters uses 1517 LUTs and 2864 FFs summarized in Table \ref{FPGA_sum}. This exhibits that the system can be implemented with minimum area and low power and hence suitable for IoT  deployable edge devices.

The resource utilization comparison contrasts related work with our system, as shown in Table \ref{RLUT}. Our work has the advantage of being low in resource utilization compared to other works. Most of these systems use acoustic signals as input, with MFCC as the feature extractor and SVM as the classification algorithm. MFCC is a widely used feature extractor for acoustic signals since it extracts linearly separable features amongst most acoustic signals. Our framework uses a neuromorphic cochlea-based kernel that acts as a feature extractor as well as a nonlinear kernel for the SVM algorithm. This avoids the need for a separate feature extractor compared to other works. Our framework also does not require separate storage for support vectors, and at the same time, we have control over the number of weights that have to be stored based on the required application. Another advantage our system has over other systems is that it is highly tunable and is scalable for higher or lower sampling frequency input signals.

\section{Results and Discussion}
We use datasets from two domains, namely speech and  environmental sounds. Speech datasets prove the usability of our framework in security like voice-based access, where we can identify the speaker and provide biometric access. The environmental sounds showcase the framework's versatility which can be deployed for multiple sounds as the target for robust classification. We use MATLAB for software simulations and verification of the algorithm. The FPGA design implements the MATLAB code using fixed-point arithmetic.

Environmental Sounds Classification (ESC-10) dataset \cite{piczak2015dataset} consists of sound clips constructed from recordings publicly available through the Freesound project. It consists of 400 environmental recordings with 10 classes, i.e., 40 clips per class and 5 seconds per clip. Each class contains 40 wav format audio files. These clips had a lot of silence, so we trimmed the silence part and further trimmed the remaining clips into 1 second version belonging to the same class, thus increasing the dataset's number of samples. Table \ref{ESC10} shows the class labels, which depict the wide variety of data samples used. The classes include sounds from dog bark, rain, sea waves, crying baby, clock ticking, person sneezing, helicopter, chainsaw, crawing rooster and fire crackling. Here, the dataset was used to create balanced classes to identify one class versus other classes arranged randomly. The train and test accuracy values are shown with the train-to-test ratio mentioned in the bracket. One thing to note from these results is that with less amount of data too, our framework could classify the sounds. We have compared our results with traditional SVM, which uses inputs after being pre-processed using the same CAR-IHC filters. For the traditional SVM, we use the in-built MATLAB library with default command lines. The number of support vectors for traditional SVMs is significantly higher than the number of filters used in our work,  indicating that we can get comparable accuracy with lower hardware resources and can be used in low-powered devices. As the number of samples is low, we see lower accuracy for classes like clock tick and person sneeze. These classes have a lot of overlapping information with other classes, causing confusion.

The Free Spoken Digit Dataset (FSDD) \cite{FSDD} is an open dataset consisting of recordings of spoken digits in wav files at a sampling rate of 8 kHz. The recordings are trimmed so that they have near minimal silence at the beginning and ends. It consists of 4 speakers with 500 recordings per speaker, amounting to an overall of 2000 recordings. These recordings are English pronunciations of each digit from 0 to 9 by each speaker. We use our framework to identify the speaker based on the recordings. We create recordings of each speaker versus a random pool of remaining speakers. We can tune our system to each speaker and get a classification to identify whether our target speaker is speaking or not. Similar to ESC dataset results, FSDD results in Table \ref{FSDD} also show that traditional SVM requires many more support vectors than the number of filters used in this work. As the number of support vectors is significantly higher for traditional SVMs, we see a slight reduction in accuracy for few classes in in-filter SVM. For the proposed in-filter SVM, the number of template vectors is determined by the fixed number of filters. The training algorithm tries to find the best possible solution within this fixed constraint. However, adhering to this constraint is one of the reasons for the reduction in accuracy. The other constraint with the proposed in-filter SVM approach is that the final solution is linear for the CAR-IHC (filter) function. Any nonlinear mapping is implemented only by the CAR-IHC function. Whereas in a standard SVM formulation that uses the CAR-IHC filter output as features, there is additional nonlinearity in the kernel mapping. Thus, the traditional SVM may be able to exploit this cross-filter nonlinearity to achieve better accuracy. FSDD classification showcases the capability of the framework to identify the right person, which can be used in giving access to a secure area or facility. 

\begin{figure}[ht]
\centerline{\includegraphics[page=1,scale=0.3,trim=0 0 0 0,clip]{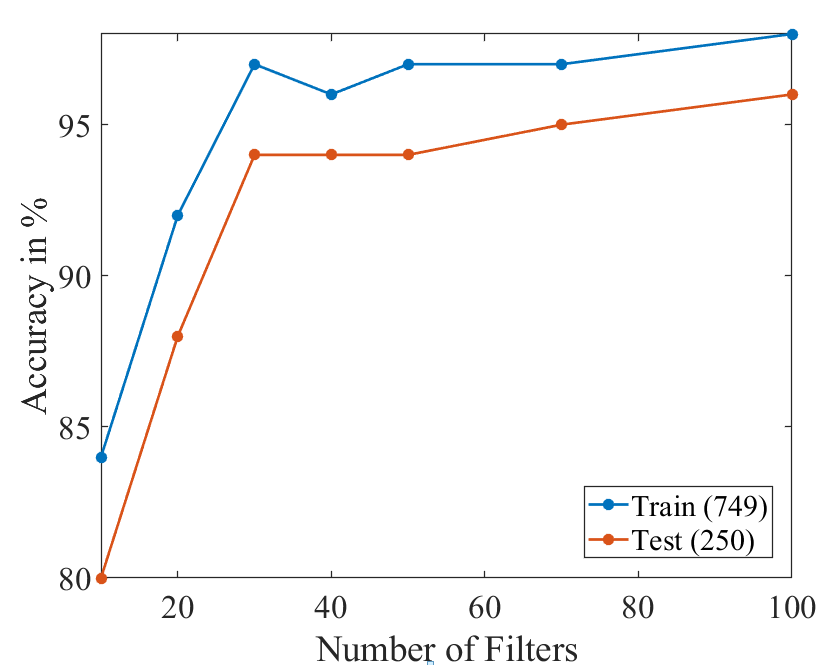}}
\caption{Impact of increasing the number of filters on accuracy for FSDD (Yweweler) dataset.}
\label{Fig:FiltersvsAcc}
\end{figure}
We see from Tables \ref{ESC10} and \ref{FSDD} that the number of support vectors ($S$) for the traditional SVM is always greater than the number of templates, i.e., filters ($P$) used for the proposed SVM ($P < S$). Traditional SVM has a computational complexity of $\mathcal{O}(S+MS)$, where $\mathcal{O}(MS)$ is the complexity of a linear kernel. In contrast, the complexity of the proposed work is $\mathcal{O}(P)$. Thus, the computational complexity of traditional SVM increases with an increase in support vectors. We see from the results that the number of filters for in-filter SVM is less than the support vectors used in traditional SVM. As a case study, we take Yweweler class data from the FSDD dataset. The number of MAC operations required to classify this class in traditional SVM is 5096 compared to 30 MAC operations for in-filter SVM. We know that MAC operations consume maximum resources and, in turn, would increase the power consumption in any hardware design. Hence, our framework is efficient in comparison to an equivalent SVM hardware implementation.

We can tune the number of filters based on the application. The number of filters is determined by the trade-off between the hardware constraints (memory and speed) versus accuracy. Empirically, we were able to determine the optimum number of filters required for most datasets. Reducing the number of filters reduces the discriminatory information encoded by the features, and hence we observe a reduction in accuracy. We can tune the number of filters based on the application. As seen in the added Fig.\ref{Fig:FiltersvsAcc}, increasing the number of filters beyond a specific value yields a marginal increase in accuracy. Thus, this marginal increase in accuracy would come at the cost of latency and increase in hardware resources, as explained in Section \ref{FPGA_Impl}. We chose 30 filters to satisfy the constraints of our implemented design. Hence, the same number of filters were used for the datasets. This shows that we can fix the number of filters based on the constraints and still obtain comparable results.

\begin{figure}[ht]
\centerline{\includegraphics[page=1,scale=0.3,trim=0 0 0 0,clip]{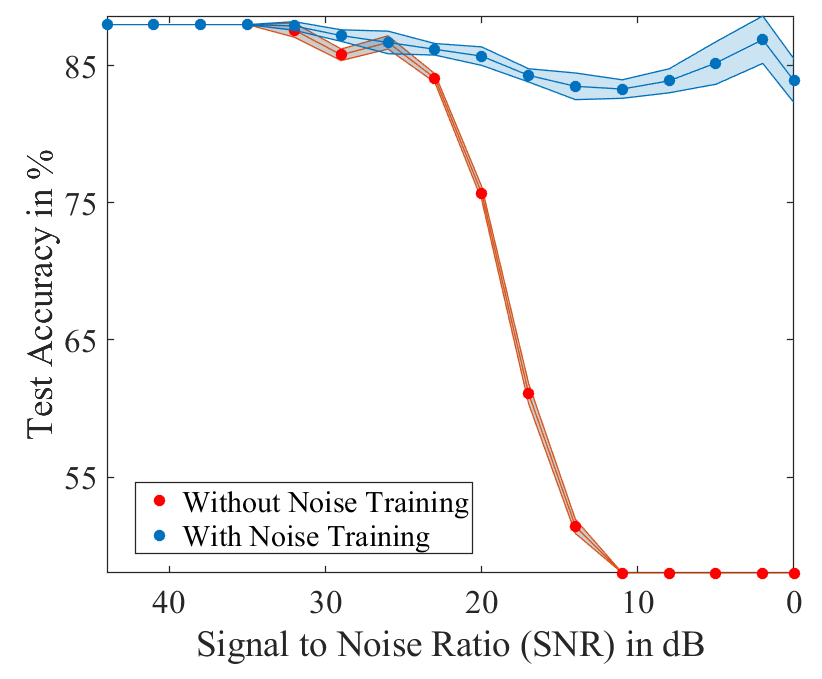}}
\caption{Impact of noise on test accuracy for FSDD (Yweweler) dataset.}
\label{Fig:SNRvAcc}
\end{figure}

We performed an experiment to check the classification robustness of our framework. For this purpose, we added white Gaussian noise to the test input signals from the existing dataset and observed the accuracies across different Signal to Noise Ratios (SNRs). We used the MATLAB tool function $awgn$ to add the white Gaussian noise to the signals. Fig.\ref{Fig:SNRvAcc} shows the mean and variance plot of the test accuracy due to the addition of noise over 10 iterations. Here, we see that our framework is quite robust when we train the data with the added noise, and with no noise in training data, the test accuracy falls below 80 \% as we reduce the SNR to below 25 dB.

\begin{table}[!htbp]
\centering
\caption{List of possible domains where our framework can be implemented by tuning the frequency ranges of the CAR filters.\label{Freq_Data}}

\begin{tabular}{|c||c|c|}
\hline
\multirow{2}{*}{\textbf{Time-series Data}} & \multicolumn{2}{c|}{\textbf{\begin{tabular}[c]{@{}c@{}}Typical Frequency \\ Range (Hz)\end{tabular}}} \\ \cline{2-3} 
                                           & \textbf{Low}          & \textbf{High}          \\ \hline \hline
Speech \cite{snow1931audible}                                     & 100                   & 8k                     \\ \hline
Music  \cite{snow1931audible}                                    & 40                    & 18k                    \\ \hline
Accelerometers   \cite{migueles2017accelerometer}                                     & 0.5                   & 1.5k                     \\ \hline
ECG   \cite{berkaya2018survey}                                     & 0.1                  & 1k                    \\ \hline
EEG  \cite{subha2010eeg}                                       & 1                   & 100                     \\ \hline
EMG \cite{komi1979emg}                                       & 24                    & 400                    \\ \hline
\end{tabular}

\end{table}

In general, we can see from the dataset results that our work produces comparable results when the number of filters is close to the number of support vectors used in traditional SVM. This shows that we can choose the number of filters beforehand and arrive at an acceptable accuracy for the required application without relying on the algorithm to decide this hardware parameter. For each type of dataset, we need to tune the filter parameters for efficient classification. We also need to determine the number of filters used, i.e., the template vectors in SVM formulation based on multiple runs. This makes our framework highly flexible and tunable as per the application's needs. In all our experiments, we have used 30 filters. The fixed point code consists of 16-bit CAR-IHC kernel output generated using 16-bit input, and the weight and bias are limited to 8-bit values. From the results across these datasets, we see that our framework is good at identifying a person using speech. Also, the ESC-10 dataset results exhibit the framework's capability even to classify inanimate sounds that can be used in systems where such classifications can trigger a more fine-tuned action for corrective measures. 
Hence, by tuning the CAR filters to a certain frequency range, we can classify different time-series data as per Table \ref{Freq_Data}. Here, our framework can be configured for a wide range of frequencies, enabling it to use various sensors generating time-series data. This gives the flexibility of programming the framework for a specific application. Also, by determining the number of filters required for each type of classification, we can optimize the classification accuracy for any time-series data.

\section{Conclusion}
In this paper, we have demonstrated our novel SVM-based acoustic classifier using the cochlea module as kernel and feature extraction stage simultaneously. The neuromorphic cochlea kernel of our unique algorithm does not require the kernel to be positive definite. This lack of restriction compared to traditional SVM enabled us to use cochlea-based CAR-IHC function as a kernel in our framework. Furthermore, the proposed system has the flexibility of handling different kinds of time-series data, as the kernel filter parameters can be tuned as per the frequency range of the input signal. This template-based SVM has a fixed number of templates in contrast to varying support vectors in traditional SVMs. We can control the operating frequency by controlling the number of kernel filters, making it power efficient. This can be fine-tuned by matching the hardware constraints with the required application speed. Also, since the complexity of this novel SVM is low compared to traditional SVM, our framework is capable of performing online training. This flexibility and dynamic behavior make the framework ideal for implementing in IoT edge devices. In this paper, we have demonstrated the hardware efficiency of the in-filter computing framework on FPGA. However, this can be extended to create a custom hardware and used as a battery-powered edge device. In the future, we plan to deploy this framework in different environments as an edge classification device. We can use this algorithm on an embedded system like a microcontroller for greater flexibility in programming the device. Leveraging the reprogramability of our framework, we can build an IoT system that can be used to monitor various time-series data using a network of sensors placed at various locations for different applications. The proposed system can have several potential applications ranging from identifying animal behaviour pattern for ecologists using sensors placed in strategic locations in a forest area to healthcare data analysis using wearable sensors which provide time-series data like ECG or EEG data. Based on bird species sounds or any animal sound, we can track the presence of different rare species of wildlife in a particular environment over a period of time. In this case, we can remotely reprogram the hardware to detect different wildlife species as many of these species might not be active in a specific region for a particular season. Similarly, such systems can also be deployed for remote health care applications using signals like ECG/EEG or ultrasound for early disease detection \cite{dogan2019optoacoustic} \cite{kang2020contrast} \cite{yu2020label} and for automation of industrial maintenance of machinery using various time-series data produced by mounted sensors. All these deployments lead to minimizing human intervention and reducing errors caused by logistics issues. Since our system can classify rare events with very low power consumption, we can deploy this system as an always-on system.

\begin{table}[hb]
\centering
\caption{List of Symbols and Acronyms.}
\begin{tabular}{|l||l|}
\hline
\textbf{Symbols and Acronyms} & \textbf{Description}                                                                                          \\ \hline \hline
$\sgn() $                     & Sign function                                                                                                 \\ \hline
$\sqrt{} $                    & Square root function                                                                                          \\ \hline
$mean() $                     & Arithmetic mean function for a series                                                                                     \\ \hline
$\underset{x}{min} f(x)$          & Find $x$ that minimizes the  function $f(x)$                                                                                    \\ \hline
$s.t.$                        & Such that                                                                                                     \\ \hline
$\mathcal{O}()$               & Big O notation for complexity                                                                                 \\ \hline
FPGA                          & Field Programmable  Gate  Array                                                                               \\ \hline
SVM                           & Support Vector Machine                                                                                        \\ \hline
CAR-IHC                       & \begin{tabular}[c]{@{}l@{}}Cascade of Asymmetric Resonator \\ with Inner Hair Cells\end{tabular}              \\ \hline
LUT                           & Look-Up  Table                                                                                                \\ \hline
FF                            & Flip-Flop                                                                                                     \\ \hline
UGS                           & Unattended   Ground   Sensor                                                                                  \\ \hline
DNN                           & Deep  Neural  Network                                                                                         \\ \hline
BNN                           & Binary Neural Network                                                                                         \\ \hline
KNN                           & K-Nearest Neighbour                                                                                           \\ \hline
M-OAA                         & Modified  One-Against-All                                                                                     \\ \hline
DWT                           & Discrete  Wavelet  Transform                                                                                  \\ \hline
RBF                           & Radial Basis Function                                                                                         \\ \hline
DSP                           & Digital Signal Processing                                                                                     \\ \hline
ROM                           & Read Only Memory                                                                                              \\ \hline
MFCC                          & Mel-frequency cepstral coefficient                                                                            \\ \hline
SRAM                          & Synchronous Random Access Memory                                                                              \\ \hline
ODE                           & Ordinary Differential Equation                                                                                \\ \hline
CARFAC                        & \begin{tabular}[c]{@{}l@{}}Cascade  of  Asymmetric  Resonators  \\ with  Fast-Acting Compression\end{tabular} \\ \hline
OHC                           & Outer Hair Cell                                                                                               \\ \hline
BM                            & Basilar  Membrane                                                                                             \\ \hline
HWR                           & Half Wave Rectifier                                                                                           \\ \hline
MSB                           & Most Significant Bit                                                                                          \\ \hline
MAC                           & Multiply-Accumulate                                                                                           \\ \hline
EEG                           & Electroencephalography                                                                                        \\ \hline
ECG                           & Electrocardiography                                                                                           \\ \hline 
ESC-10                        & Environmental Sound Clips-10                                                                                  \\ \hline
FSDD                          & Free Spoken Digit Dataset
                                        \\ \hline
EMG                           & Electromyography                                                                                           \\ \hline
SNR                           & Signal to Noise Ratio                                                                                \\ \hline
\end{tabular}
\end{table}

\appendix[CAR-IHC Filter Formulation]

A two pole two zero filter forms the asymmetric resonator whose transfer function is as below:
\begin{eqnarray}
    H(z) = \frac{Y}{X} = g[\frac{(z-z_{zero})(z-z^{*}_{zero})}{(z-z_{pole})(z-z^{*}_{pole})}].  \nonumber \\
     = g[\frac{z^{2}+(-2a_{0}+kc_{0})rz+r^{2}}{z^{2}-2a_{0}rz+r^{2}}].
\end{eqnarray}
The two pole coupled form has a pair of conjugate poles ($z_{pole}$ and $z^{*}_{pole}$):
\begin{eqnarray}
    z_{pole},z^{*}_{pole} = \frac{2a_{0}\pm \sqrt{(2a_{0}r)^{2}-4r^{2}}}{2}.\\
    = rcos(\theta_{R})\pm irsin(\theta_{R}).\nonumber
\end{eqnarray}
\begin{equation}
    a_{0} = cos(\theta_{R}).
\end{equation}
where $\theta_{R}$ is the pole angle in the z plane. The conjugate zeros ($z_{zero}$ and $z^{*}_{zero}$) are:
\begin{equation}
  z_{zero},z^{*}_{zero} = \frac{-(-2a_{0}+kc_{0})r}{2}. \nonumber \\
   \pm \frac{ \sqrt{((-2a_{0}+kc_{0})r)^{2}-4r^{2}}}{2}. 
\end{equation}
\begin{equation}
    = rcos(\theta_{Z})\pm irsin(\theta_{Z}). \nonumber 
\end{equation}
\begin{equation}
    a_{0}-\frac{kc_{0}}{2} = cos(\theta_{Z}).
\end{equation}
where $\theta_{Z}$ is the zero angle in the z plane. The zero radius is the same as the pole radius, $r$. The condition for complex zeros becomes relevant for high-frequency channels, where $cos(\theta_{R}) < 0$:
\begin{equation}
    a_{0}-\frac{kc_{0}}{2} > -1.
\end{equation}
\begin{equation}
    k < \frac{2+2a_{0}}{c_{0}}.
\end{equation}
Here, $r$ can be used to move the zeros and the poles simultaneously while $k$ is fixed. $k$ determines the distance between the poles and the zeros. The frequency of zeros are kept slightly higher than the poles.  If we increase the value of $k$, the poles and zeros grow further apart, giving a slow roll off at higher frequencies. On the other hand, if the value of $k$ is decreased to a low value, the poles and zeros grow closer, giving rise to sharp roll off making it asymmetric. This sharp roll off is similar to the characteristic exhibited by auditory filtering. This property also enhances selection of frequencies. In order to keep the pole frequency half octave below zero frequency, $k$ is kept sames as $c_{0}$.
\IEEEPARstart{} To get unity gain at DC, we can solve for g as follows:
\begin{equation}
    g= \frac{1 - 2a_0r+r^2}{1 - (2a_0 - k c_0)r+r^2}.
\end{equation}

\IEEEPARstart{} The zero-crossing times of the filter’s impulse
response does not change with respect to time, even when we change $r$. 
\begin{equation}
    r = 1 - damping \times \frac{2 \pi f}{f_s}.
\end{equation}
where $damping$ controls the damping factor, $f$ is defined in eq.\eqref{eq:27} and $f_s$ is the sampling frequency. $r$ keeps the damping away from zero and also makes the damping bounded.
Changing $r$ means varying the poles and the zeros of the filter. This satisfies the biologically observed condition where variation in stimulus level does not vary the impulse response zero crossings \cite{lyon2017human}. For each cascade stage, the initial values for zeros and poles are set. The Greenwood map function  \cite{greenwood1990cochlear} is used to choose equidistant poles of the two pole two zero resonator. These are placed along the normalized length of the cochlea.

\begin{equation}
    f = 165.4(10^{2.1x}-1). \label{eq:27}
\end{equation}
where, $f$ is the frequency of the pole and $x$ is the normalized position along the cochlea,
varying from 0 at the apex of the BM, to 1 at the basal end.


%



\section*{Acknowledgment}
This research was supported in part by (i) INSPIRE faculty fellowship 
(DST/INSPIRE/04/2016/000216), SPARC grant (SPARC/2018-2019/P606/SL) from Ministry of Human Resource Development and IMPRINT Grant IMP/2018/000550 from the Department of Science 
and Technology, India. 
The authors would like to acknowledge the joint Memorandum of Understanding (MoU) between Indian Institute of Science, Bangalore and Washington University in St. Louis for supporting this research activity.

\ifCLASSOPTIONcaptionsoff
  \newpage
\fi



%
\medskip


\bibliographystyle{IEEEtran}
\bibliography{references}

\end{document}